\documentclass[aps,prL,twocolumn,superscriptaddress,showpacs,floatfix]{revtex4-1}
\usepackage{amsmath,amssymb,graphicx}
\usepackage{wasysym}
\usepackage[utf8]{inputenc}
\usepackage[T1]{fontenc}
\usepackage{xcolor}
\allowdisplaybreaks
\usepackage{rotating} 
\usepackage[columnwise]{lineno}
\usepackage{verbatim}

\usepackage{textcomp}

\usepackage{bm}

\IfFileExists{siunitx.sty}{\usepackage{booktabs,siunitx}}{}

\usepackage{color}
\definecolor{LinkColor}{rgb}{0.256,0.439,0.588}
\usepackage{hyperref}
\hypersetup{
    pdfstartview={FitH},    
    colorlinks=true,       
    linkcolor=blue,          
    citecolor=blue,        
    filecolor=magenta,      
    urlcolor=blue           
}

\usepackage{pifont}

\newcommand{\La}{\line (1,0  ){5}}      

\newcommand{\Ld}{\line (-2,0){5}}     

\newcommand{\C} {\circle{3}}        

\newcommand{\LaT}{\rule[-0.7pt]{0.18cm}{0.15em}}  
\newcommand{\LdT}{\rule[-0.7pt]{0.18cm}{0.15em}}  
\newcommand{\LbT}{\rotatebox{60}{\rule[-1pt]{0.18cm}{0.15em}}}  
\newcommand{\LeT}{\rotatebox{60}{\rule[-1pt]{0.18cm}{0.15em}}}  

\newcommand{\LbTs}{\rotatebox{60}{\rule[1.2pt]{0.18cm}{0.05em}}}  
\newcommand{\LeTs}{\rotatebox{60}{\rule[2.3pt]{0.18cm}{0.05em}}}  

\newcommand{\pZ}{\put(-0.5,0)}    
\newcommand{\pC}{\put(7.5,0)}   
\newcommand{\pA}{\put(-4,-7)}    
\newcommand{\pB}{\put(4,-7)} 

\newcommand{\pZb}{\put(0.8,0)}    
\newcommand{\pCb}{\put(6,0)}   
\newcommand{\pAb}{\put(-2.5,-7)}    
\newcommand{\pBb}{\put(4.3,-6.5)} 

\newcommand{\pAT}{\put(-4.3,-5.5)} 
\newcommand{\pBT}{\put(3.7,-5.5)}  

\newcommand{\rhomb}{            
  \pA{\C}\pB{\C}\pZ{\C}\pC{\C}
 }

\newcommand{\rhombH}{            
  \begin{picture}(20,5)(-8,-6)
    \pAb{\LaT}\pBb{\LbTs}\pA{\LeTs}\pZb{\LdT}
    \rhomb
  \end{picture}
}
\newcommand{\rhombV}{            
  \begin{picture}(20,5)(-8,-6)      
   \pAb{\La}\pBT{\LbT}\pAT{\LeT}\pCb{\Ld}
    \rhomb
  \end{picture}
}

\newcommand{\hdimer}{{\hrule height0.05cm width0.2cm depth0pt}}
\newcommand{\mdimer}{\vbox{\hdimer \vskip 0.0625cm}}

\newcommand{\ohordimer}{\hbox{\vbox{\mdimer}}}

\newcommand{\vdimerr}{\rotatebox{330}{{\vrule height0.2cm width0.05cm depth0pt}}}
\newcommand{\overdimerr}{\hbox{\hskip 0.05cm \vdimerr }}

\newcommand{\vdimerrr}{\rotatebox{30}{{\vrule height0.2cm width0.05cm depth0pt}}}
\newcommand{\overdimerrr}{\hbox{\hskip 0.05cm \vdimerrr }}

\begin{document}

\title{Phase transitions and remnants of fractionalization at finite temperature in the triangular lattice quantum loop model}

\author{Xiaoxue Ran}
\affiliation{Department of Physics and HK Institute of Quantum Science \& Technology, The University of Hong Kong, Pokfulam Road, Hong Kong SAR, China}

\author{Sylvain Capponi}
\affiliation{Laboratoire de Physique Th\'eorique, Universit\'e de Toulouse, CNRS, UPS, France}

\author{Junchen Rong}
\affiliation{Institut des Hautes \'Etudes Scientifiques, 91440 Bures-sur-Yvette, France}
\affiliation{CPHT, CNRS, \'Ecole Polytechnique, Institut Polytechnique de Paris, Palaiseau, France}

\author{Fabien Alet}
\affiliation{Laboratoire de Physique Th\'eorique, Universit\'e de Toulouse, CNRS, UPS, France}

\author{Zi Yang Meng}
\email{zymeng@hku.hk}
\affiliation{Department of Physics and HK Institute of Quantum Science \& Technology, The University of Hong Kong, Pokfulam Road, Hong Kong SAR, China}

\begin{abstract}
The quantum loop and dimer models are archetypal correlated systems with local constraints. With natural foundations in statistical mechanics, they are of direct relevance to various important physical concepts and systems, such as topological order, lattice gauge theories, geometric frustrations, or more recently Rydberg arrays quantum simulators. However, how the thermal fluctuations interact with constraints has not been explored in the important class of non-bipartite geometries. Here we study, via unbiased quantum Monte Carlo simulations and field theoretical analysis, the finite-temperature phase diagram of the quantum loop model on the triangular lattice. We discover that the recently identified, "hidden" vison plaquette (VP) quantum crystal~\cite{ran2024hidden} experiences a finite-temperature continuous transition, which smoothly connects to the (2+1)d Cubic* quantum critical point separating the VP and $\mathbb{Z}_{2}$ quantum spin liquid phases. This finite-temperature phase transition acquires a unique property of ``remnants of fractionalization" at finite temperature, in that, both the cubic order parameter -- the plaquette loop resonance -- and its constituent -- the vison field -- exhibit independent criticality signatures. This phase transition is connected to a 3-state Potts transition between the lattice nematic phase and the high-temperature disordered phase. We discuss the relevance of our results for current experiments on quantum simulation platforms.
\end{abstract}

\maketitle

\noindent{\textcolor{blue}{\it Introduction.}---} Quantum dimer/loop models (QDM/QLM) serve as quintessential statistical models to describe the low-energy properties of many-body systems with local constraints occurring in frustrated magnets~\cite{Moessner2001l,moessnerShort2001,Krishanu2015,Plat2015,Yan2021,ZYan2022,yanEmergent2023,Moessner2001b,Ivanov2004,Ralko2005,Misguich2008,Dabholkar2022Reentrance,ranFully2023} or cold atom experiments ~\cite{Bernien17,keesling2019quantum,PhysRevLett.124.103601,Ebadi.2021,scholl2021quantum,ZYan2022,PhysRevB.105.174417,Samajdar:2020hsw,Verresen:2020dmk}. Given that each site on the lattice can only be connected to one/two dimer/loop segments, QDM/QLM encompass a wide range of phenomena such as 
spinless and topological excitations in unconventional superconductors~\cite{Kivelson1987,Rokhsar1988,Baskaran1988,sachdev2021translational}, and facilitate the detection of new phases and dynamic phenomena in frustrated systems~\cite{Moessner2000Two,Moessner2001l,Jiang2005String,ran2024hidden,Ran2024Cubic}. The mapping between the QDM/QLM and Rydberg atom arrays further enables the observation of emergent phases in programmable quantum simulators~\cite{Bernien17,Samajdar:2020hsw,Semeghini21,Giudici2022Dynamical,ZYan2022,zhou2022u1,yanEmergent2023}.

The triangular lattice quantum loop model 
\begin{eqnarray}
  H=&-t&\sum_\alpha \left(
  \left|\rhombV\right>\left<\rhombH\right| + \mathrm{h.c.}
  \right) \nonumber \\
  &+V&\sum_\alpha\left(
  \left|\rhombV\right>\left<\rhombV\right|+\left|\rhombH\right>\left<\rhombH\right|
  \right),
\label{eq:eq1}
\end{eqnarray}
where $\alpha$ denotes all the rhombi (with three orientations) on the triangular lattice, is a perfect illustration of how highly constrained systems on non-bipartite geometries can give rise to a wide variety of physical phenomena. For the QLM, the local constraint requires that every site is touched by two loop segments in any configuration. The kinetic term $t$ (set to $1$ in the following) generates plaquette loop resonance while respecting the local constraint, and $V>$ is the repulsion ($V > 0$) or attraction ($V < 0$) between loop segments on resonating plaquettes. Its ground-state phase diagram has been investigated intensively, notwithstanding a few controversies~\cite{Krishanu2015,Plat2015,ran2024hidden}. At the Rokshar-Kivelson point $V=t$~\cite{rokhsarSuperconductivity1988}, it is agreed that a $\mathbb{Z}_2$ quantum spin liquid (QSL) phase is formed, which subsists down to weaker repulsion $V_c<1$. This phase of matter (observed in Rydberg arrays~\cite{Semeghini21}) hosts fractionalized quasiparticles known as visons. Reducing further the strength of repulsion, the condensation of visons gives rise to different symmetry-breaking crystals~\cite{Moessner2001b,Ivanov2004,Ralko2005,Misguich2008,Yan2021}. Initial studies~\cite{Krishanu2015,Plat2015} indicated that
the crystal is a lattice nematic (LN) phase with a QSL-LN transition suggested~\cite{Krishanu2015} to belong to the 3D O(3)* universality class, with * meaning that the transition is towards a topologically non-trivial $\mathbb{Z}_2$ QSL with fractionalized excitations~\cite{sachdev_book} and differs from a 3D O(3) transition as diagnosed by large anomalous dimensions. Analogous anomalous phenomena due to fractionalization appear in the XY* transition~\cite{Isakov2012,YCWang2018,sachdev_book} which separates a topologically trivial ordered phase from a $\mathbb{Z}_2$ QSL.

However, due to the difficulties in studying quantum constrained models in an unbiased manner, the results presented in these works disagree strongly on the value of $V_c$. Further, the speculation that the LN–QSL transition belongs to the O(3)* universality class was based on the assumption that the cubic anisotropy (due to the microscopic lattice symmetries) is irrelevant at the
3D O(3) fixed point~\cite{aharony1973critical,hasenbuschAnisotropic2011,Adzhemyan:2019gvv,Aharony:2022ajv,Pelissetto:2000ek}, but this has been recently disproven by conformal bootstrap analyses showing that the renormalization group instead flows to the nearby corner cubic fixed point~\cite{Chester2021,rong2023o3}.
Very recently, Ref.~\onlinecite{ran2024hidden} identified, with the help of an improved quantum Monte Carlo (QMC) method, a new "hidden" vison plaquette (VP) phase located between LN and $\mathbb{Z}_{2}$ QSL phases. At odds with the LN where the $C_3$ lattice rotational symmetry-breaking can be identified from loop patterns, the VP phase is symmetric in the loop configurations but exhibits translational symmetry-breaking in the off-diagonal loop resonance term (the kinetic term in Eq.~\eqref{eq:eq1}) and its hidden constituent -- the fractionalized vison field. The quantum critical point (QCP), separating the VP and $\mathbb{Z}_{2}$ QSL was found to be of the O(3) Cubic* universality class~\cite{ran2024hidden,Ran2024Cubic} whereas the quantum phase transition between the LN and VP phases is first order~\cite{ran2024hidden}. 

The occurrence of this novel VP phase and the quantum critical behavior out of it, raises the intriguing questions on its {\it finite-temperature} ($T$) melting. The interplay between quantum and thermal fluctuations in the vicinity of a QCP~\cite{Sachdev_2011} can offer further finite-$T$ signatures and insights of quantum criticality~\cite{Paiva2005Ground,Yael2010Finite,Xu2017Non,YDL2021,huang2024interaction}. Almost all previous works focused on ground state phases of QLM/QDM~\cite{Moessner2001l,moessnerShort2001,Moessner2001b,Ralko2005,Ralko2006,Syljuasen2006Plaquette,Misguich2008,Sikora2011Extended,Krishanu2015,Plat2015,Yan2021,ZYan2022,ran2024hidden,Ran2024Cubic}, with the only exception (to the best of our knowledge) of the investigation of the finite-$T$ phase diagram of the square lattice QDM~\cite{Dabholkar2022Reentrance}, exhibiting Kosterlitz-Thouless transition and high-$T$ criticality intrinsically connected to the bipartiteness of the square lattice. The critical behavior on {\it non}-bipartite lattices is unexplored at nonzero $T$, and is of particular interest as their $T=0$ phase diagram generically differs from their bipartite counterparts (as illustrated by the existence of the VP phase for the triangular QLM).

\begin{figure}[htp!]
\centering
\includegraphics[width=\linewidth]{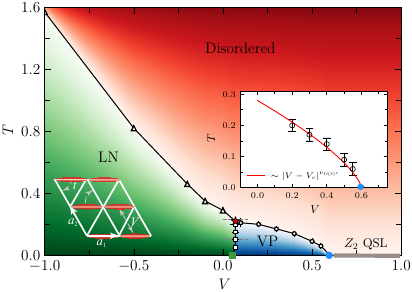}
\caption{Finite-temperature phase diagram of the QLM on the triangular lattice which exhibits different phases: Lattice Nematic (LN, green region), Vison Plaquette (VP, blue region), and disordered (red) phases. A comprehensive summary of the data can be found in Table. S1 and S3 of the Supplemental Material (SM)~\cite{suppl}. The solid gray horizontal line marks the $\mathbb{Z}_{2}$ spin liquid phase only existing in the ground-state. The green $(V \simeq 0.05)$ and blue $(V\simeq 0.6)$ points correspond to the LN-VP and VP-QSL quantum phase transitions. The red star indicates a possible multicritical point at $(V, T)=(0.07(2), 0.22(2))$. The two dashed gray lines correspond to $T=0.23$ (upper) and $T=0.1$ (lower) used in Fig.~\ref{fig:fig4}. The inset zooms in at low-$T$ near the VP-QSL Cubic* QCP, highlighting that the finite-$T$ phase boundary scales as $T_c(V) \sim |V-V_{c}|^{\nu_{C^*}}$, with $\nu_{C^*}\simeq 0.7$ the correlation length exponent for the 3d Cubic* universality class.}
\label{fig:fig1}
\end{figure}

In this  work, we investigate the finite-$T$ phase diagram of the QLM on the triangular lattice addressing large systems with the sweeping cluster QMC method~\cite{Yan2019, ZY2020improved} combined with a field-theoretical analysis. Our results are summarized in the phase diagram of Fig.~\ref{fig:fig1}. As the temperature is increased, we observe a 3-state Potts continuous phase transition between the LN phase and the disordered phase. More interestingly, we discover that the VP crystal experiences a finite-$T$ continuous transition, which smoothly connects to the 3d Cubic* QCP separating the VP and QSL phases. This transition exhibits ``remnants of the fractionalization" observed at the QCP in that, both the plaquette loop resonance (the cubic order parameter) and the vison fields (its constituent) exhibit criticality with a symmetry structure corresponding to the rank-2 symmetric traceless tensor and rank-1 fundamental vector representations of the Cubic* criticality. Although strong finite-size effects do not allow us to make definitive statements about the exponents of this VP-Disorder continuous transition, we discuss interesting possible scenarios. This transition with ``remnants of fractionalization" at finite temperature connects to the 3-state Potts transition line above the LN phase, as well as to a first-order phase transition line between the LN and VP phases at low $T$. 

\noindent{\textcolor{blue}{\it Simulations and Order Parameters.}---}
We study Eq.~\ref{eq:eq1} using QMC simulations for system sizes $N=L^2$ with $L = 8, 12, 16, 20, 24, 28, 32$.  On top of lattice symmetries, the model Eq.~\eqref{eq:eq1} exhibits topological sectors labeled by the parity of the winding number of loops. We restrict our QMC simulations in the $(0,0)$ topological sector to decrease statistical errors, but we checked using a fully-ergodic extension of the sweeping cluster algorithm~\cite{Dabholkar2022Reentrance} as well as exact diagonalization data (see SM~\cite{suppl}) that this does not affect our conclusions as the finite-$T$ properties discussed below are dominated by the $(0,0)$ sector. 

In the field theoretical description of the Cubic* VP-QSL QCP~\cite{ran2024hidden}, the Lagrangian couples three $(i=1,2,3)$ scalars $\phi_i$ representing the vison modes~\cite{Krishanu2015}:
\begin{eqnarray}
\label{cubicstar}
\mathcal{L}_{int}=m^2(\sum_i \phi_i^2)+u(\sum_i \phi_i^2)^2 + \nu(\sum_i \phi_i^4) + \cdots \\
 \label{eq:eq3}
{\rm with} \quad \phi_i = \sum_{\mathbf{r}}(v_1(\mathbf{r}),v_2(\mathbf{r}))\cdot \mathbf{u}_j e^{i\mathbf{M}_j\cdot\mathbf{r}},
\end{eqnarray}
where $\mathbf{r}=(x, y)$ is the coordinate of the visons located at the center of the triangle. The $\mathbf{M}_{i=1,2,3}$ are the 3 $\mathbf{M}$ points of the Brillouin zone that minimize the energy dispersion, and $\mathbf{u}_j $ are the eigenvectors at these momenta with $u_1=u_2=(1,1)^T$ and $u_3=(1,-1)^T$. The vector $(\phi_1, \phi_2, \phi_3)$ encapsulates the 3d Cubic order parameter of the visons. The action was derived in Ref.~\cite{Krishanu2015} as possibly describing the O(3)* LN-QSL QCP (not present in the triangular QLM). Identifying the mass term as $m^2 \sim |V - V_c|$, the VP-QSL phase transition occurs when $m^2 = 0$, and was shown using QMC to belong to the 3d Cubic* universality class~\cite{Ran2024Cubic} with critical exponents in agreement with estimates from Monte Carlo simulations and conformal bootstrap \cite{ballesterosFinite1996,hasenbuschAnisotropic2011,Chester2021,hasenbuschCubic2023}. 
   
The order in the VP phase can also be identified from the static structure factor $C{\cal T}_{\mathbf{q}=\mathbf{M}}$  of the real space $t$-term correlator $\langle {\cal T}_{i} {\cal T}_{j} \rangle$ at the $\mathbf{M}$ points 
\begin{equation}
C{\cal T}_{\mathbf{q}=\mathbf{M}}=\frac{1}{3L^2}\sum_{l=1}^{3}\sum_{j=1}^{L^2}\sum_{k=1}^{L^2}e^{-i\textbf{M}_{l}\cdot(\mathbf{R}_j-\mathbf{R}_k)}\langle {\cal T}_{j}{\cal T}_{k}\rangle.
\end{equation}
Here $l$ averages over three $\mathbf{M}$ points, $\mathbf{R}$ denotes the coordinates of all upper triangles and ${\cal T}_j=t_{1,j}+t_{2,j}+t_{3,j}$ the sum of the kinetic $t$-terms over the three rhombi centered around an upper triangle $j$. In the symmetry description of the Cubic* criticality~\cite{ran2024hidden,Ran2024Cubic}, the kinetic terms $(t_{1}=\phi_1\phi_2, t_{2}=\phi_2\phi_3, t_{3}=-\phi_1\phi_3)$ form the rank-2 symmetric traceless tensor representation of the Cubic group, while the vison modes $(\phi_1, \phi_2, \phi_3)$ form the rank-1 fundamental vector representation. To detect criticality, it is useful to consider the correlation ratio $R_{C{\cal T}}=1-C{\cal T}_{\mathbf{q'}}/C{\cal T}_{\mathbf{q}=\mathbf{M}}$, where $C{\cal T}_{\mathbf{q'}}$ is the average structure factor at the momenta $1/L$ away from $\mathbf{M}$. 

Finally, the nematic order parameter of the LN phase is $\langle D \rangle = \frac{1}{3L^{2}}\sum_{c}(|N^c_{\overdimerr}-N^c_{\ohordimer}|+|N^c_{\overdimerrr}-N^c_{\ohordimer}|
+|N^c_{\overdimerr}-N^c_{\overdimerrr}|)$,
where $N^c_{\overdimerr}$, $N^c_{\overdimerrr}$, and $N^c_{\ohordimer}$ count loop segments in configuration $c$ for the three distinct orientations. 
We now discuss the finite-$T$ transitions out of the LN and VP phases based on these order parameters.

\begin{figure}[htp!]
	\centering
	\includegraphics[width=\linewidth]{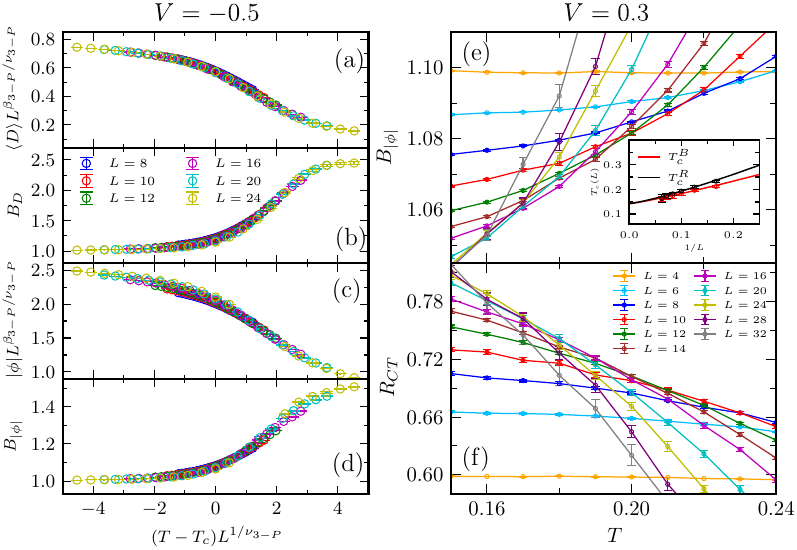}
	\caption{Left panels: data collapse for (a) the nematic order parameter $\langle D \rangle$, (b) its Binder ratio $B_{D}$, (c) the vison parameter $\phi$ and (d) its Binder ratio $B_{|\phi|}$ for system sizes from $L=8$ to $L=24$. Simulations are performed at $V=-0.5$, where the ground state is deep within the LN phase. All data collapses for $\langle D \rangle$, $B_{D}$, $\phi$ and $B_{|\phi|}$ use the critical exponents $\beta_{3P}=\frac{1}{9}$ and $\nu_{3P}=\frac{5}{6}$ of the 3-state Potts model. Right panels: (e) Binder ratio $B_{|\phi|}$ of the vison order parameter and (f) Correlation ratio $R_{C{\cal T}}$ of the $t$-terms for system sizes $L=4$ to $L=32$ for $V=0.3$ (where the ground state is located within the VP phase). The inset illustrates the crossing between system sizes $L$ and $2L$ for both $B_{|\phi|}$ and $R_{C{\cal T}}$. Lines are fits to $T_{c}(L)=T_{c}(\infty)+aL^{-(\omega+1/\nu)}$, where $\omega$ is the correction exponent and $\nu$ the correlation length critical exponent. For $V=0.3$, we obtain $T_{c}(\infty)=0.14(1)$ and $\omega+1/\nu\sim1.4$. Similar analysis for other $V$ values are presented in SM~\cite{suppl}.}
	\label{fig:fig2}
\end{figure}

\begin{figure}[htp!]
	\centering
	\includegraphics[width=\linewidth]{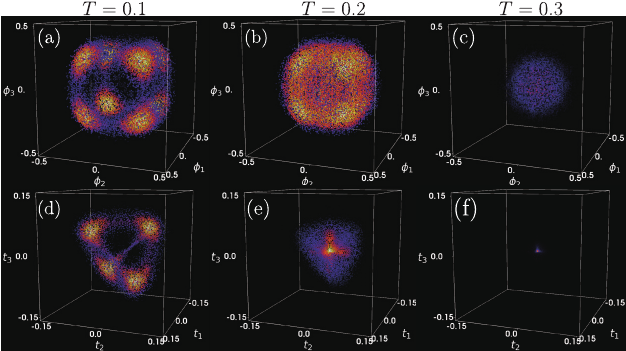}
	\caption{Three-dimensional histograms of ($\phi_{1}, \phi_{2}, \phi_{3}$) (first row) and $(t_{1}, t_{2}, t_{3})$ (second row) within the VP phase ($T=0.1$), near the transition point ($T=0.2$) and in the disordered phase ($T=0.3$). System size is $L=24$.  They illustrate the phase transition process, where the corner-cubic order in the VP phase, the cube in ($\phi_{1}, \phi_{2}, \phi_{3}$) and tetrahedron shapes in $(t_{1}, t_{2}, t_{3})$, shrink to a point in the disordered phase.}
	\label{fig:fig3}
\end{figure}

\noindent{\textcolor{blue}{\it LN-Disorder 3-State Potts  Transition.}---}
Starting on the left part of the phase diagram, we obtain a very good finite-size collapse of the nematic order parameter $\langle D \rangle$ and its Binder ratio $B_{D}=\langle D^4 \rangle / \langle D^2 \rangle^2$ in Fig.~\ref{fig:fig2} (a) and (b) at $V=-0.5$ using the critical exponents of the 3-state Potts universality class $\beta_{3P}=\frac{1}{9}$ and $\nu_{3P} =\frac{5}{6}$~\cite{baxter2007exactly,wu1982potts,ALEXANDER1975353,baxter2007exactly,wu1982potts}, with the interpretation that the three states in the Potts model correspond to the possible orientations of loops in the LN phase. The empty triangles in  Fig.~\ref{fig:fig1} denote the LN-Disorder transitions at other values of $V$ as determined from a similar analysis.
In addition to the nematic order parameter, we also monitor the critical behavior of the vison modes $\phi$ for the same LN-Disorder transition. The data collapse of $|\phi|$ and its Binder ratio $B_{|\phi|}={\langle|\bm{\phi}|^4\rangle}/{\langle|\bm{\phi}|^2\rangle^2}$ with $|\phi|=\sqrt{\phi_{1}^2+\phi_{2}^2+\phi_{3}^2}$ at $V=-0.5$ are shown in Fig.~\ref{fig:fig2} (c) and (d) with the same exponents as for the nematic order parameter. This can be intuitively understood as the existence of a symmetry-breaking pattern in the loop configurations (leading to a finite expectation value for $\langle D \rangle$) induces in a deterministic fashion a similar symmetry-breaking  for the $\phi$ field (see SM~\cite{suppl} for a schematic presentation of $\phi$ and loop-segment fields) 
with both 3-fold rotation and translation being broken. Therefore $D$ or $|\phi|$ are the same order parameter for the LN-Disorder transition, and there is no remnant of fractionalization at finite-$T$ in this part of the phase diagram.

\noindent{\textcolor{blue}{\it VP-Disorder Transition and ``remnants of fractionalization" at finite-$T$.}---}
The transition out of the intrinsically quantum VP phase is more subtle as it cannot be easily captured by a classical constrained model with the same symmetries (as e.g. a classical dimer model captures the finite-$T$ columnar transition in the square lattice QDM~\cite{Dabholkar2022Reentrance}). 
We first discuss theoretical expectations for the VP-Disorder transition before comparing to numerical simulations. Near the QCP at $V_c \simeq 0.6$ of Eq.~\eqref{cubicstar}, the critical temperature should follow a power law behavior $T_c\sim |V-V_c|^{\nu_{C^*}}$, with $\nu_{C^*}$ the critical exponent of the Cubic* QCP (for more details see SM~\cite{suppl}). The finite-temperature VP-Disorder transition has two possible interpretations. 
Putting Eq.~\eqref{cubicstar} at finite $T$, we expect a scalar theory respecting the Cubic symmetry in two spatial dimensions. One candidate theory is three decoupled copies of Ising models since the Cubic conformal field theory (CFT) merge with the decoupled Ising models at $D=2$. Away from the quantum critical point, however, another candidate is the 4-state Potts universality class. In this case, the gauge invariant order parameter for the VP phase is given by the kinetic $t_i$-terms, which becomes the order parameter of the 4-state Potts model. 
The details of the above discussion are explained in SM~\cite{suppl}. The crucial difference between these two interpretations are whether the kinetic $t_i$-terms and the vison field $\phi$ have different scaling dimensions at criticality.

\begin{figure}[htp!]
	\centering
	\includegraphics[width=\linewidth]{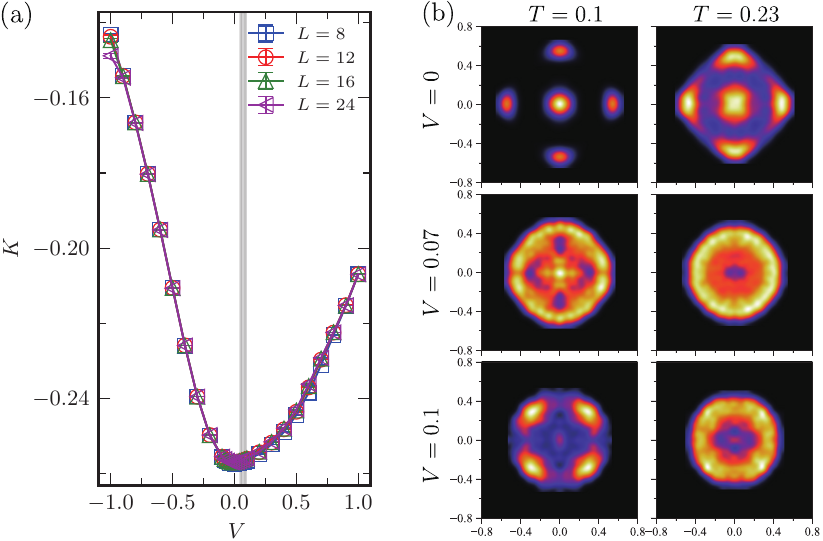}
	\caption{(a) Kinetic energy density $K=\sum_j {\cal}\langle T_j \rangle/L^2$ for $L=8$ to $L=24$ as a function of $V$ at $T=0.1$. The vertical line indicates the transition at $V=0.07(2)$. (b) Histograms of $\phi$ at $T=0.1$ and $T=0.23$ (dashed gray lines in Fig.~\ref{fig:fig1}) for $L=24$, with three rows corresponding to $V=0$, $V=0.07$, $V=0.1$. The disappearance of the VP phase at all values of $V$ as the temperature increases to $0.23$ suggests the LN phase directly transitions to a disordered phase. }
	\label{fig:fig4}
\end{figure}

To test this numerically, we first follow in Fig.~\ref{fig:fig3} both kinetic ${\cal T}$-term and vison field $\phi$ histograms across the VP-Disorder transition , to determine the transition temperature. In the VP phase, the three $\phi_i$ take values in the eight vertices of a cube while the scalar bilinear triplet $t_i$  take four different values, forming a tetrahedron (see Fig.~\ref{fig:fig3} (a), (d)). Further, Fig.~\ref{fig:fig2} (e), (f) depict the Binder  $B_{|\phi|}$ and correlation $R_{C{\cal T}}$ ratios at $V=0.3$ (in the middle of the VP phase) as a function of $T$ for system sizes $L$ up to 32. Both quantities exhibit crossings at finite $T$ -- a signature of a continuous VP-Disorder transition, albeit with obvious drifts of the crossing point with $L$. Performing a finite-size scaling analysis~\cite{Luck1985Corrections, qinDuality2017,chenPhase2024} allows to determine the transition temperature in the thermodynamic limit (see inset of the right panels of Fig.~\ref{fig:fig2}), and critical exponents. We obtain $T_{c}(\infty)=0.14(1)$ and $\omega+1/\nu\sim1.4$ (the $T_c$ at other $V$ values and the corresponding exponents are listed in SM~\cite{suppl}). While resulting in consistent determinations of $T_c$ (open circles in Fig.~\ref{fig:fig1}), the data analysis in the SM suggests that the critical exponents do not take a constant value along the thermal VP-disorder phase transition, and moreover that the estimated exponents are quite far off the candidate universality classes. We cannot exclude that this is due to the limited sample sizes simulated, and that on larger length scales, the exponents crossover --possibly slowly due to log corrections-- to those e.g. of the 4-state Potts or decoupled Ising universality class, see~\cite{suppl}. It is also possible that finite-size estimates are affected by the proximity to the QCP or to a possible multicritical point where LN, VP and disordered phases meet (the red star in Fig.~\ref{fig:fig1}). 
Even though we can not determine the critical exponents accurately, it is quite remarkable that the $\phi$ and ${\cal T}$-terms appear to have different exponents as shown in Table.~\ref{tablebeta}, a situation quite different from the LN-Disorder transition but mimicking the QCP Cubic* behavior~\cite{ran2024hidden}.
This could be interpreted in favor of the decoupled Ising scenario, instead of the 4-state Potts one. We furthermore plot the finite-$T$ phase boundary $T_c(V)$ of the VP phase in the inset of Fig.~\ref{fig:fig1}, which is is well-fitted by $T_c(V) \sim  |V-V_c|^{\nu_{C^*}}$, where $V_c=0.6$ and $\nu_{C^*}\simeq 0.7$~\cite{Ran2024Cubic}. Taken together, these two facts suggest that the VP-Disorder thermal transition is partially controlled by the Cubic* QCP at finite $T$, at least in its vicinity and on the finite sizes accessible to us. 

\begin{table}[h]
    \centering
    \caption{Critical exponent obtained from the finite-size scaling analysis on $\phi$ and $C{\cal T}$ data.}
    \begin{tabular}{cccc} 
        \hline
           $V$& $0.2$ &$0.3$  &$0.4$   
            \\ \hline
        $\beta$ from $\phi$ & $0.189(6)$   & $0.22(2)$ & $0.212(8)$    \\ 
        $\beta$ from $C{\cal T}$  & $0.44(1)$ & $0.528(7)$ & $0.45(2)$\\
        \hline
    \end{tabular}
    \label{tablebeta}
\end{table}

\noindent{\textcolor{blue}{\it LN-VP Transition and Multicritical point.}---} The LN-VP transition at $T=0$ was identified to be first order based on the phase coexistence in the $\phi$ histograms~\cite{ran2024hidden} that change from face-cubic to corner-cubic shapes. We display these  histograms for the $L=24$ system for finite $T=0.1$ and $T=0.23$ in Fig.~\ref{fig:fig4} (b). A coexistence of peaks located on face centers (for LN) and corners (for VP) of the cube indicates a first-order phase transition at $V=0.07$, which is further confirmed by the observed change in the slope of the kinetic energy density in Fig.~\ref{fig:fig4}(a). The LN-VP phase boundary (white squares in Fig.~\ref{fig:fig1}) is approximately vertical. The VP phase vanishes around $T=0.23$, as shown in the lower right corner of Fig.~\ref{fig:fig4}(b). This suggests the existence of a multicritical point $(V, T)=(0.07(2), 0.22(2))$ marked by a red star in Fig.~\ref{fig:fig1} (see SM~\cite{suppl} for further supporting data).

\noindent{\textcolor{blue}{\it Discussion.}---}
We obtained the finite-temperature phase diagram the triangular lattice QLM, via unbiased large-scale QMC assisted by a field-theoretical approach. We discovered that the VP crystal experiences a finite-$T$ continuous transition, with signatures of ``remnants of fractionalization" at finite temperature, in that, both the cubic order parameter (the plaquette loop resonance) and its constituent (the vison field) exhibit critical behavior with symmetry structure corresponding to the rank-2 symmetric traceless tensor and rank-1 fundamental vector representations of the Cubic* criticality. This structure is maintained inside the entire VP phase, and the finite-$T$ transition smoothly connects to the 3d Cubic* QCP separating the VP and $\mathbb{Z}_{2}$ QSL phases, and to a 3-state Potts transition above the LN phase via a multicritical point.  

We finally discuss relevance of our finite-$T$ phase diagram for current experiments on quantum simulators. Symmetry breaking at finite $T$ has already been observed in 2d Rydberg arrays~\cite{chen_continuous_2023}, where the finite $T$ is effectively obtained through an excess energy induced by a variable quench introduced in an adiabatic ramp protocol. Another possibility to emulate finite-$T$ states is to use short-time quantum dynamics measurements on quantum simulators and retrieve thermal results using hybrid classical-quantum algorithms~\cite{lu_algorithms_2021}. The effectiveness of this appraoch has been demonstrated numerically~\cite{schuckert_probing_2023} and experimentally~\cite{hemery_measuring_2024}. Further, experimental advances in creating dual-species Rydberg arrays~\cite{singh_dual_2022,anand_dual_2024} allow for the potential preparation of finite-temperature states via sympathetic cooling~\cite{raghunandan_initialization_2020}. Regarding the required measurements to diagnoze fractionalization, the LN or vison correlators can be directly experimentally imaged in Rydberg arrays (being diagonal in the loop configuration/Rydberg state occupation basis) while kinetic term (off-diagonal) correlators will require global rotation prior to the detection, similar to the measurement protocol in Ref.~\cite{chen_continuous_2023}. Our results thus provide strong motivations to look for such phenomena in experiments on Rydberg atom arrays or other synthetic platforms.

\vspace{0.5cm}
\noindent{\it Acknowledgments.----} We acknowledge discussions with Bhupen Dabholkar, Yang Qi and Zheng Yan. We acknowledge the support from the
Research Grants Council (RGC) of Hong Kong (Project Nos. AoE/P701/20, 17309822, C7037-22GF, 17302223, 17301924), the HKU Seed Funding for Strategic Interdisciplinary Research (X.R.R. and Z.Y.M.) and the ANR/RGC Joint Research Scheme sponsored by RGC of Hong Kong (Project No. A\_HKU703/22) and French National Research Agency (grant ANR-22-CE30-0042-01) (X.R.R., S.C., F.A. and Z.Y.M.). 
J.R. acknowledges the funding from the Simons programme g\'en\'eral (2022-2031) in Institut des Hautes \'Etudes Scientifiques. 
J.R. also acknowledges funding and from the European Union (ERC “QFTinAdS”, project number 101087025). 
Views and opinions expressed are however those of the authors only and do not necessarily reflect those of the European Union or the European Research Council Executive Agency. Neither the European Union nor the granting authority can be held responsible for them.
We thank HPC2021 system under the Information Technology Services, University of Hong Kong, CALMIP (grant 2024-P0677), GENCI (project A0150500225) as well as the Beijing PARATERA
Tech CO.,Ltd. (URL: https://cloud.paratera.com) for providing HPC resources that have contributed to the research results reported within this paper.

\bibliographystyle{longapsrev4-2}
\bibliography{main}
\setcounter{equation}{0}
\setcounter{figure}{0}
\setcounter{table}{0}
\renewcommand{\theequation}{S\arabic{equation}}
\renewcommand{\thefigure}{S\arabic{figure}}
\renewcommand{\thetable}{S\arabic{table}}

\clearpage
\newpage
\setcounter{page}{1}
\begin{widetext}
\linespread{1.05}

\centerline{\bf Supplemental Material for}

\centerline{\bf "Phase transitions and remnants of fractionalization at finite temperature}
\centerline{\bf in the triangular lattice quantum loop model"} 
\vskip3mm

\centerline{}
In this {\color{black}Supplemental Material}, we first present the review of the mapping between the dimer and vison configurations and the ground-state phase diagram for the QLM on the triangular lattice in Sec.~\ref{sec:gs}. Then, the results of exact diagonalization for QLM on the triangular lattice are shown in Sec.~\ref{sec:ED}. In Sec.~\ref{sec:svp}, we discuss finite-size scaling and data collapse through a Bayesian scaling analysis of the order parameters in the VP-Disorder phase transition. Thereafter, the determination of the LN-VP transition boundary and the estimation of the multicritical point are performed through an analysis of the histograms of $\phi$ in Sec.~\ref{sec:first}. Supplemental QMC data for the LN-disordered phase transition is presented in Sec.~\ref{sec:ln}. Finally, Sec.~\ref{sec:SMV} offers a field-theoretical discussion on the finite-temperature VP-Disorder transitions.

\section{Ground-state phase diagram for the QLM on the triangular lattice}
\label{sec:gs}
In this section, we review the mapping between the loop and vison configurations and the ground-state phase diagram for the QLM on the triangular lattice, as shown in Fig.~\ref{fig:gs} and obtained in Ref.~\cite{ran2024hidden}.

The vison configuration $v_{\gamma}(\mathbf{r})$ ($\gamma=1,2$) is obtained from a loop configuration by setting a reference value $v_{1}(0,0)=1$ for the upper triangle of the first unit cell, as illustrated in Fig.~\ref{fig:gs}(a). Two visons located at the centers of related triangular plaquettes $i$ and $j$ can be connected with a open string, as depicted by the green dashed line in Fig.~\ref{fig:gs}(a). Their correlation is given by $\langle v_i v_j\rangle=\langle(-1)^{N_{P_{ij}}} \rangle$, where ${N_{P_{ij}}}$ is the number of loop segments intersected along a path $P$ between plaquettes $i$ and $j$. This correlation reflects the topological nature of the vison excitations and their dependence on the underlying loop configuration in the system.

In Fig.~\ref{fig:gs}(b), we display the ground-state phase diagram of the QLM on the triangular lattice~\cite{ran2024hidden}. We show three possible LN loop configurations in the left subfigure, while the corresponding vison patterns are indicated by the green values assigned to each triangle. A first-order phase transition between the LN and VP phases occurs at $V=0.05(5)$. In the middle subfigure, we present a schematic representation of the real-space vison correlation functions as obtained in Ref.~\cite{ran2024hidden}. The green (grey) color denotes the positive (negative) correlation values, while the depth of the colors represents the relative magnitude of these values. The right subfigures illustrate representative dimer coverings in the $\mathbb{Z}_{2}$ QSL phase. A continuous phase transition between the VP phase and the QSL occurs at {$V_{c}=0.59(2)$}.

\begin{figure}[htp!]
	\centering
	\includegraphics[width=1.0\textwidth]{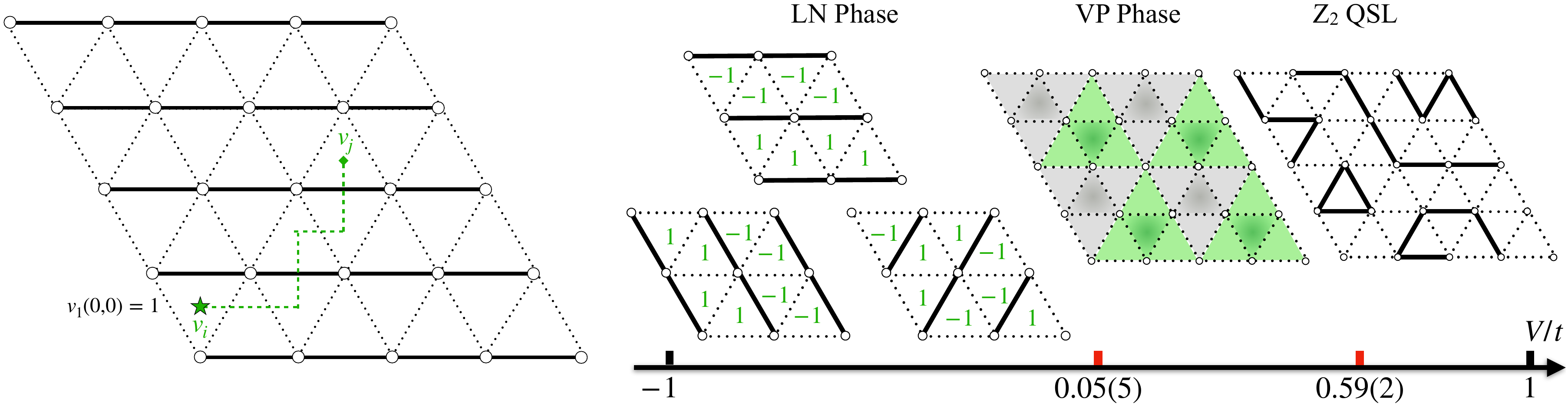}
	\caption{(a) The QLM and the string operator between two visons on a triangular lattice. The open circles represent the lattice sites, while solid black bonds indicate the presence of loop segments on those bonds. Dashed bonds represents the absence of loop segments. The reference vison is marked by a green star at the upper triangle of the first unit cell. Two visons $v_i$ and $v_j$, located on the triangles, are connected by a string, represented by the dashed green line. (b) Ground-state phase diagram of the QLM on the triangular lattice as obtained in Ref.~\cite{ran2024hidden}. In the left panel, we display three possible loop configurations within the LN phase, with the corresponding vison patterns indicated by the green numbers in each triangle. In the middle subfigure, we present a schematic representation of the real-space vison correlation functions. The color scheme employs green to denote positive correlation values and grey for negative values, while the intensity of the colors reflects the relative magnitudes of these correlations.}
	\label{fig:gs}
\end{figure}

\section{Exact diagonalization results}
\label{sec:ED}

In order to get some insight and to benchmark QMC results, we performed exact diagonalization on small clusters using periodic boundary conditions of the model Eq. (1) in the main text. On a tilted 28-site cluster (the largest cluster we could consider), there are more than 7 billion loop configurations but this number can be reduced by using various symmetries: 
(i) first, we can work in a given topological sector (TS), one trivial and three degenerate nontrivial ones; (ii) second, we can use all translations to reduce the Hilbert space size and also obtain the momentum; (iii) third, we can use the full point-group symmetry (here $C_6$ for this cluster) depending on the momentum value. Overall this allows to reduce the size to at most 66 millions configurations. We have used the Lanczos algorithm to compute the low-energy spectrum. In Fig.~\ref{fig:figED_28}, we present our results as a function of $V$. 
For large negative $V$, one can clearly identify  three quasi-degenerate states in the trivial TS, all at $\Gamma$ momentum, as expected for the LN phase that only breaks rotation symmetry in the thermodynamic limit. 
Close to the RK point ($V=1$), we observe quasi degenerate ground states in each TS, one trivial and three degenerate in the nontrivial TSs, as expected for the QSL. In between, there is an additional low-energy state with momentum M in the trivial TS, as expected for the VP phase.

\begin{figure}[htp!]
	\centering
	\includegraphics[width=0.6\textwidth]{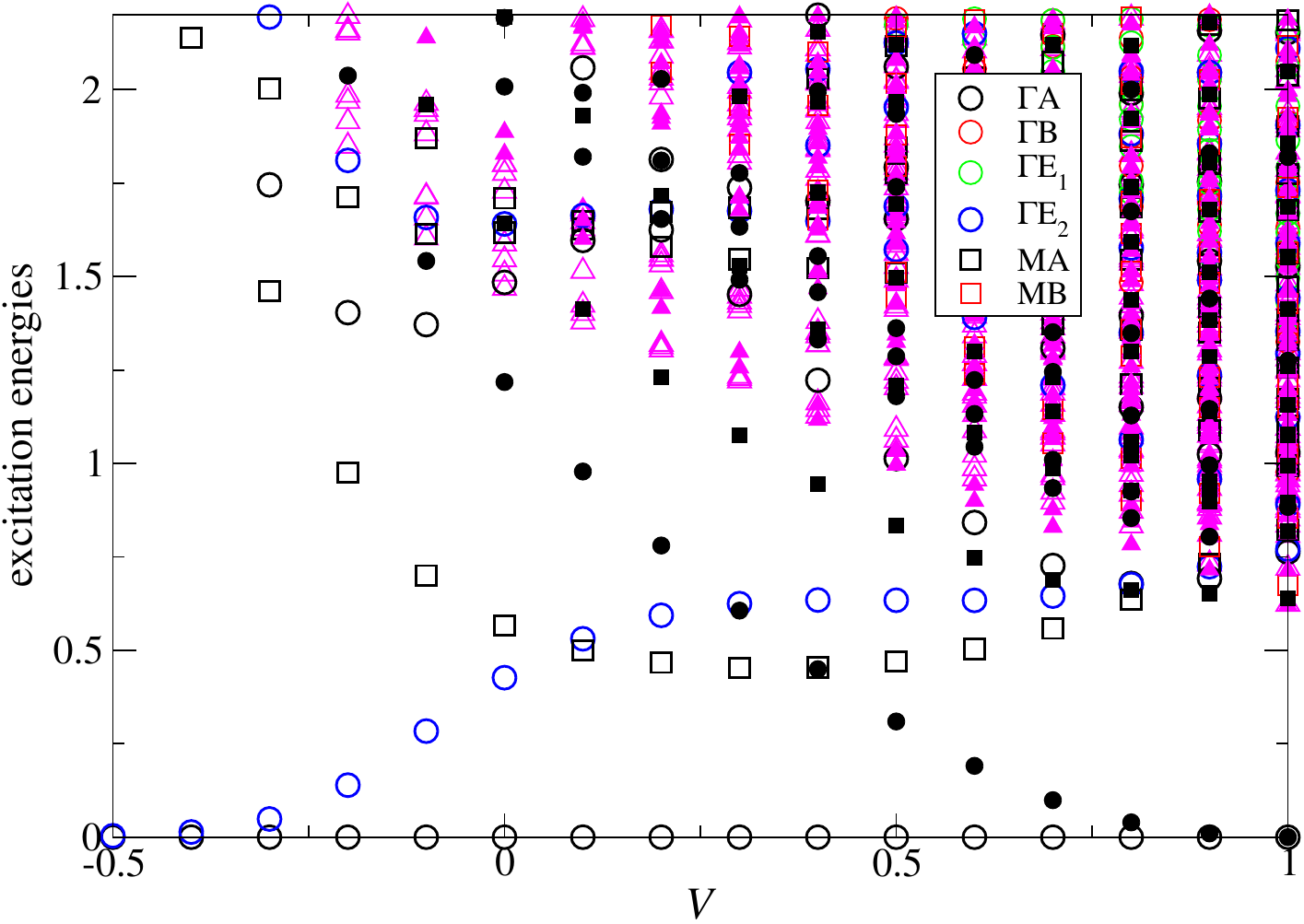}
	\caption{Low-energy excitations versus $V$ obtained from ED on $N=28$ sites cluster. This cluster has $C_6$ point group symmetry and does  contain the M point. The energy levels are labelled with their momenta, point-group symmetry as well as the topological sector (trivial/nontrivial are shown respectively with open/filled symbols). We highlight some relevant energy levels at momentum $\Gamma$ (labelled using the usual $C_6$ irreps notations) as well as momentum M (labelled using A/B for even/odd states with respect to the reflection symmetry).}
	\label{fig:figED_28}
\end{figure}

Quite importantly, we also note that the topological gap, i.e. minimum excitation energy in the nontrivial TSs, is quite large for most values of the interaction $V$ that we have considered in the QMC simulations. This justifies that QMC simulations can be performed solely in the trivial TS to capture the universal properties of the finite-temperature LN or VP melting transitions. In particular, all the critical temperatures found in the QMC simulations are  below this topological gap.

\section{Finite-size scaling analysis and data collapse for the VP-Disorder phase transition}
\label{sec:svp}
In this section, we investigate the criticality of the phase transition between the VP phase and the disordered phase using finite-size scaling and data collapse techniques. This analysis includes the determination of the transition point and aims to identify the associated universality class.

For dimensional quantities A near the critical point, including only the leading correction term, the scaling function can be expressed as~\cite{Luck1985Corrections} 
\begin{equation}
A(T, L)=L^{\kappa/\nu}[f^{(0)}(tL^{1/\nu})+L^{-\omega}f^{(1)}(tL^{1/\nu})]
\end{equation}
where $t=T-T_{\text{c}}$ is the reduced temperature, and $\omega$ is the correction exponent. For systems of $L$ and $\lambda L$ (we use $\lambda=2$ in this study), the Taylor expansion of $f$'s suggests that the crossing points of curves of the Binder and correlation ratios (both dimensionless quantities, that is $\kappa=0$) as a function of $T$ satisfy the relation ~\cite{Derrida1988275,Luck1985Corrections,Shao2016Quantum,ma2018anomalous,Ran2019}
\begin{equation}
T_{c}(L)=T_{c}(\infty)+aL^{-(\omega+1/\nu)}
\end{equation}
where $a$ is a fitting parameter. We perform a stochastic extrapolation for the crossing points $T_{c}(L)$ as a function of $1/L$ for both $B_{|\phi|}$ and $R_{C{\cal T}}$ to determine the transition temperature in the thermodynamic limit. Firstly, to locate the crossings of two system sizes, we fit the Binder ratio and correlation ratio using a third order polynomial form (a quartic polynomial yields similar results). In the nonlinear fitting with the scaling form $T_{c}(L)=T_{c}(\infty)+aL^{-(\omega+1/\nu)}$, we initially set ranges for each parameter and and randomly select their values to minimize $\chi^2/d.o.f$. This selection process is repeated extensively ($\sim 10^5$ times) until each parameter converges. We expect all converged parameters to exhibit a Gaussian distribution for optimal fits with $\chi^2/d.o.f\sim 1$. In Fig.~\ref{fig:dis}, we show the scaling results at $V=0.3$ with $L\geq6$ as an example, similar results are observed for other values of $V$. The extrapolation results for a few selected values of $V$ are presented in the first column of Table~\ref{table1}. The extrapolation for $V=0.3$ is illustrated in the inset of Fig. 2 in the main text with the weighted average of converged parameters. We found that the transition temperature determined by this extrapolation has a relatively small value and a larger error bars than those from data collapses in Table~\ref{table1}. A possible explanation is that this is due the possible residual effects from further subleading correction terms, which are not well extracted given our current data quality.

\begin{figure}[htp!]
	\centering
	\includegraphics[width=0.6\linewidth]{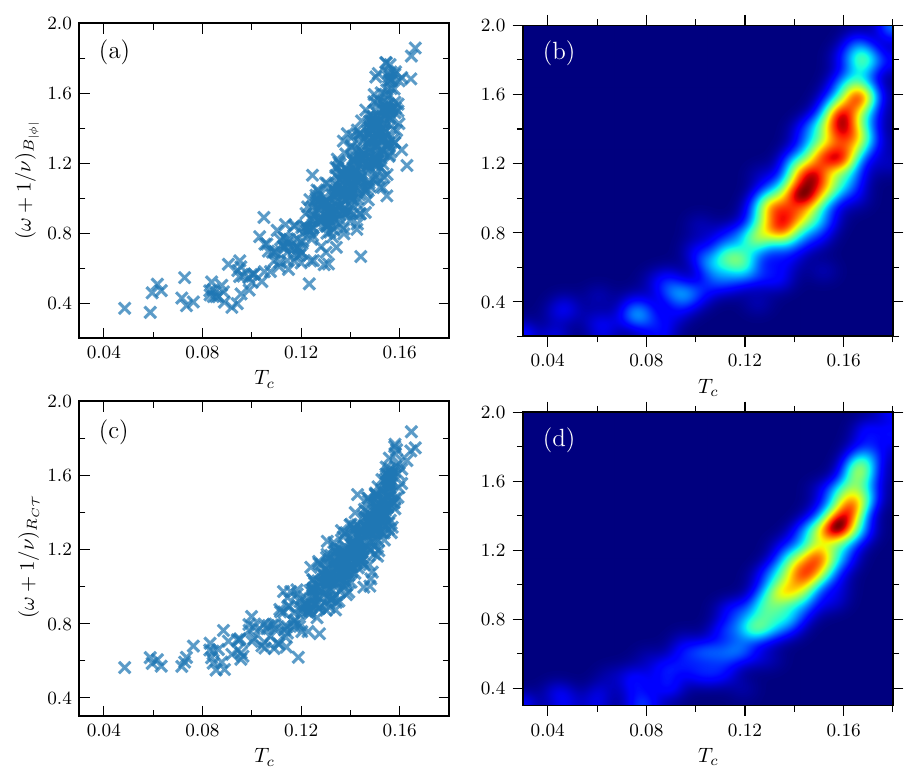}
	\caption{Finite-size scaling results for the crossings from Binder ratio and correlation ratio at $V=0.3$. The crossings from two observables are fitted together with scaling forms $T_{c}(L)_{B_{|\phi|}}=T_{c}(\infty)+aL^{-(\omega+1/\nu)_{B_{|\phi|}}}$ and $T_{c}(L)_{R_{C{\cal T}}}=T_{c}(\infty)+bL^{-(\omega+1/\nu)_{R_{C{\cal T}}}}$. In the left column, we show the converged distributions of the transition point $T_c$ and critical exponents. Each data point represent a fit with $\chi^2/d.o.f\sim 1$. The right panels display their heat map to show the density of the data points in the left panels. We use the weighted average of parameters obtained in this stochastic extrapolation process to show the extrapolation in the inset of Fig. 2 in the main text. }
	\label{fig:dis}
\end{figure}

The data collapse for order parameters and their Binder ratio $B$, defined as
\begin{align}
|\phi|(T, L)L^{\beta^{\phi}/\nu}&=tL^{1/\nu},\\
C{\cal T}(T, L)L^{2\beta^{C{\cal T}}/\nu}&=tL^{1/\nu},\\
B(T, L)&=tL^{1/\nu},
\end{align}
and presented in panels (b), (c), (e), and (f) of Fig.~\ref{fig:figs1}, are conducted using a method based on Bayesian scaling analysis, as introduced in Ref~\cite{Kenji2011Bayesian,Kenji2015Kernel,GPlink}. This approach does not require an explicit fitting form and yields superior scaling results compared to traditional least-squares methods for our data.

\begin{figure}[htp!]
	\centering
	\includegraphics[width=\linewidth]{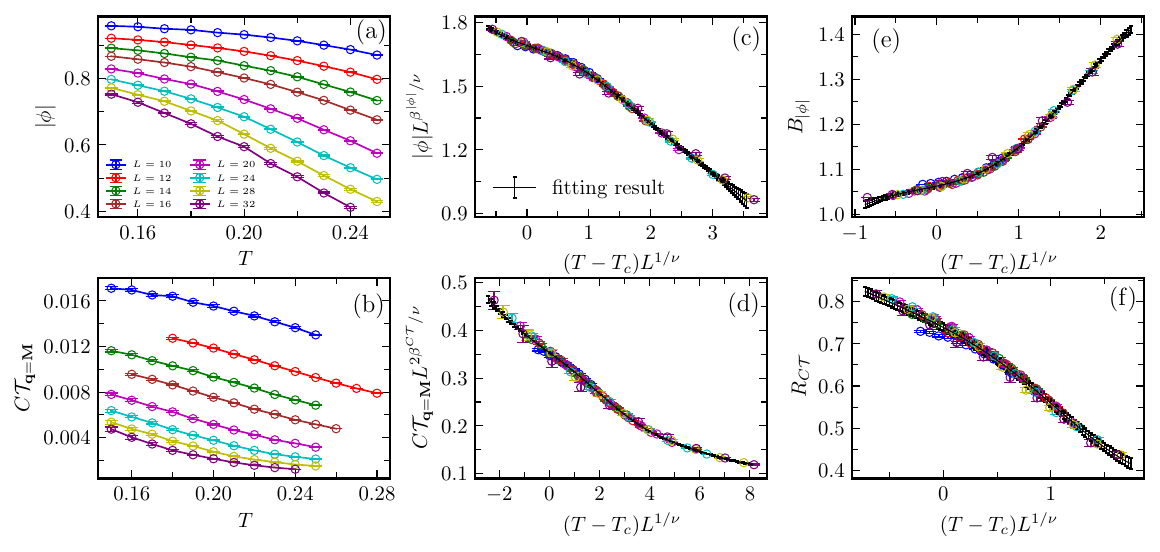}
	\caption{Order parameters and their data collapse at $V=0.3$. The first column displays (a) the vison order parameter $|\phi|$ and (b) the static structure factor at the $M$ momenta of the $t$-term correlations $C{\cal T}$. The middle column shows the data collapse for these quantities using Bayesian scaling analysis in panels (c) and (d). The right column presents (e) the data collapse of the Binder ratio of $|\phi|$ and (f) the correlation ratio of $C{\cal T}$, with raw data provided in Fig. 3 of the main text. The fitting results for the panels (c), (d), (e), and (f) are summarized in Table~\ref{table1}.}
	\label{fig:figs1}
\end{figure}

The fit results for different values of $V$ within the VP phase are summarized in Table.~\ref{table1}. We note that for values of $V$ close to the ground-state phase transition points, scaling becomes increasingly challenging. For instance, the data collapse for $C{\cal T}$ at $V=0.1$ fails unless we fix $T_c=0.2$, as indicated in Table.~\ref{table1}. similarly, the data collapses for $C{\cal T}$ and $R_{C{\cal T}}$ also fail at $V=0.55$. For each $V$, we computed a weighted average of $T_c$ from the data collapse results of different physical observables, using the Log-likelihood from each data collapse result as the weight, as indicated in Table.~\ref{table1}. The weighted average is calculated using $\sum(w_{i}x_{i})/\sum w_{i}$, with the errorbar given by $\sqrt{\sum(w_{i}^2\sigma_{i}^2)/(\sum w_{i})^2}$. The final estimated $T_c$ in the last column of Table.~\ref{table1} is chosen to balance all estimates, with an errorbar encompassing all uncertainties. The larger error bar in Fig. 1 of the main text reflects the combined uncertainty from all methods, thereby reducing bias toward any single scaling.
 
When comparing these results with the critical exponents of the two-dimensional Ising model, 4-state Potts model and related Ashkin-Teller (AT) model listed in Table.~\ref{table2}, we first observe that our values for $\beta^{\phi}$ and $\nu$ lie within the range of critical exponents characteristic of the AT model.  In the AT model, while the critical exponents $\beta$ and $\nu$ vary along the critical line, their ratio remains fixed. However, our data yield a ratio of $\beta^{\phi}$ and $\nu$ of approximately $1/4$, which differs from the $1/8$ ratio of the AT model. Furthermore, our critical exponents derived from the data collapse differ substantially from those of the 4-state Potts model or Ising model. On the other hand, we also observe that the values of $2 \beta^{C{\cal T}}/\nu$ (obtained from the data collapse for $C{\cal T}$ and which corresponding to the scaling dimension of the kinetic ${\cal T}$ term) are quite large $\sim 1.4$ for $V=0.2,0.3$ and $0.4$, and close to the scaling dimension of the rank-2 symmetric traceless tensor for the 3d Cubic* and O(3)* transition. This seems to suggest that the observed finite-temperature phase boundary of VP-Disorder transition, still experiences the influence of the ground state VP-QSL 3d Cubic* QCP. It is important to note that due to the limitations in system size in our simulations, the quality of the extrapolation and data collapse are not optimal.

\begin{table}[h]
    \centering
    \caption{Fitting results for order parameters in the VP phase. The merged column shows the data collapse results for different observables.}
    \begin{tabular}{c|c|ccccc|c} 
        \hline
        $V=0.1$ & Extrapolation  & $|\phi|$ & $B_{|\phi|}$  & $C{\cal T}$ & $R_{C{\cal T}}$ & Weighted average & Estimated $T_c$ \\ \hline
        $T_{c}$                  & 0.190(8)       & 0.226(2)     & 0.247(3) &  Fixed 0.2        & 0.254(2) & 0.238(2) &0.21(3)\\ 
        $\beta^{\phi}$           &               &  0.212(6)   &               &              &            &  &   \\ 
        $\beta^{C{\cal T}}$             &               &             &   &   0.59(2)                  &           &  &     \\ 
        $\nu$                    &               & 0.92(2)     & 0.80(6)  &   0.97(3)        & 0.92(6) &0.89(4)  &\\ 
        $(\omega+1/\nu)_{B_{|\phi|}}$           & 1.1(2)        &              &               &              &   &  &\\
         $(\omega+1/\nu)_{R_{C{\cal T}}}$           & 1.1(1)        &              &               &              &       &  &  \\ 
          Log-likelihood          &   &  250.8            &   137.4          &    163.8       &  108.2      &     &  \\ \hline
          
          $V=0.2$ &  &  &  &  &  &  &\\ \hline
         $T_{c}$                  & 0.176(7)       & 0.199(2)     & 0.206(2)      & 0.204(1)     & 0.219(3) & 0.206(2) &0.20(2)\\ 
        $\beta^{\phi}$           &               & 0.189(6)     &               &              &            &  &   \\ 
        $\beta^{C{\cal T}}$             &               &              &               & 0.44(1)      &           &  &     \\ 
        $\nu$                    &               & 0.79(2)      & 0.90(4)       & 0.67(2)      & 0.90(2) &0.79(3)  &\\ 
        $(\omega+1/\nu)_{B_{|\phi|}}$           & 1.4(2)        &              &               &              &   &  &\\
         $(\omega+1/\nu)_{R_{C{\cal T}}}$           & 1.3(1)        &              &               &              &       &  &  \\ 
          Log-likelihood          &   &  153.4            &    138.4          &     233.9         &  104.5      &     &  \\ \hline
        
        $V=0.3$ &  &  &  &  &  &  &\\ \hline
        $T_{c}$              & 0.14(1)       & 0.161(4)      & 0.168(2)       & 0.177(1)       & 0.181(2) & 0.172(3) & 0.17(2)\\ 
        $\beta^{\phi}$       &               & 0.22(2)       &                &                &            &  &   \\ 
        $\beta^{C{\cal T}}$         &               &               &                & 0.528(7)       &            &  &    \\ 
        $\nu$                &               & 0.91(4)       & 1.01(4)        & 0.741(8)       & 0.98(2) &0.89(3)  &\\ 
        $(\omega+1/\nu)_{B_{|\phi|}}$           & 1.2(2)        &              &               &              &  &  & \\
        $(\omega+1/\nu)_{R_{C{\cal T}}}$           & 1.2(2)        &              &               &              &        &  & \\ 
        Log-likelihood        &        & 166.7             & 179.1             & 249.1            & 144.7 &  &      \\ \hline

        $V=0.4$ &  &  &  &  & &  & \\ \hline
        $T_{c}$              & 0.112(8)       & 0.135(2)     & 0.142(2)       & 0.138(1)         & 0.150(2) & 0.140(2) & 0.14(2)\\ 
        $\beta^{\phi}$       &               & 0.212(8)     &                &                  &          &  &     \\ 
        $\beta^{C{\cal T}}$         &               &              &                & 0.45(2)          &           &  &     \\ 
        $\nu$                &               & 0.77(3)      & 0.82(4)        & 0.63(2)          &0.68(1) & 0.72(3) &\\ 
        $(\omega+1/\nu)_{B_{|\phi|}}$           & 1.1(3)        &              &               &              &  &  & \\
        $(\omega+1/\nu)_{R_{C{\cal T}}}$           & 1.4(2)        &              &               &              &      &  &   \\
        Log-likelihood          &   &   208           &      173.6        &     217.2         &   120.2     &     & \\ \hline
        
                $V=0.5$ &  &  &  &  & &  & \\ \hline
        $T_{c}$              & 0.093(6)       & 0.098(3)     & 0.107(2)       & 0.097(3)         & 0.111(2) & 0.102(3) & 0.09(2)\\ 
        $\beta^{\phi}$       &               & 0.23(2)     &                &                  &          &  &     \\ 
        $\beta^{C{\cal T}}$         &               &              &                & 0.50(2)          &           &  &     \\ 
        $\nu$                &               & 0.73(4)      & 0.77(4)        & 0.64(3)          &0.66(4) & 0.70(4) &\\ 
        $(\omega+1/\nu)_{B_{|\phi|}}$           & 1.3(3)        &              &               &              &  &  & \\
        $(\omega+1/\nu)_{R_{C{\cal T}}}$           & 1.9(2)        &              &               &              &      &  &   \\
        Log-likelihood          &   &   144.9          &      121.3        &    151.1        &   102.8     &     & \\ \hline
        
                $V=0.55$ &  &  &  &  & &  & \\ \hline
        $T_{c}$              & 0.060(9)       & 0.048(5) & 0.087(2)           &          &  & 0.067(4) & 0.06(2)\\ 
        $\beta^{\phi}$       &               & 0.36(2)     &                &                  &          &  &     \\ 
        $\beta^{C{\cal T}}$         &               &              &                &         &           &  &     \\ 
        $\nu$                &               & 1.01(5)      & 0.63(3)        &        & &0.83(4)   &\\ 
        $(\omega+1/\nu)_{B_{|\phi|}}$           &         &              &               &              &  &  & \\
        $(\omega+1/\nu)_{R_{C{\cal T}}}$           & 2.0(3)        &              &               &              &      &  &   \\
        Log-likelihood          &   &   146.7           &      132.7        &             &       &     & \\ \hline
        
    \end{tabular}
    \label{table1}
\end{table}

\begin{table}[h]
    \centering
    \caption{Critical exponents for two-dimentional Ashkin-Teller (AT) models on the critical line, 4-state Potts model, and Ising model~\cite{Kohmoto1981Hamiltonian,baxter2007exactly}}
    \begin{tabular}{cccc} 
        \hline
            \hspace{0.5cm} &$\beta$\hspace{0.5cm}  &$\nu$ \hspace{0.5cm} &$\beta/\nu$ \hspace{0.5cm}      
            \\ \hline
        \\
        AT\hspace{0.5cm}   & $\left[\frac{1}{12}, \frac{1}{4}\right]$   \hspace{0.5cm}                    &$\left[\frac{2}{3}, 2\right]$ \hspace{0.5cm} &$\frac{1}{8}$ \hspace{0.5cm}       \\ 
        \\
        4-state Potts \hspace{0.5cm}  &$\frac{1}{12}$\hspace{0.5cm} &$\frac{2}{3}$\hspace{0.5cm} &$\frac{1}{8}$ \hspace{0.5cm} \\
        \\
        Ising \hspace{0.5cm}  &$\frac{1}{8}$\hspace{0.5cm} &$1$\hspace{0.5cm} &$\frac{1}{8}$ \hspace{0.5cm} \\
        \\ \hline
    \end{tabular}
    \label{table2}
\end{table}

\begin{figure}[htp!]
	\centering
	\includegraphics[width=0.9\linewidth]{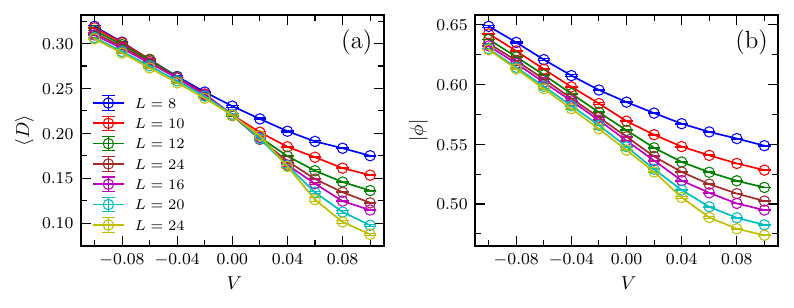}
	\caption{Order parameters (a) $\langle D \rangle$ and (b) $|\phi|$ as a function of $V$ at $T=0.1$ for various system sizes.}
	\label{fig:figs4}
\end{figure}

\begin{figure}[htp!]
\centering
\includegraphics[width=\linewidth]{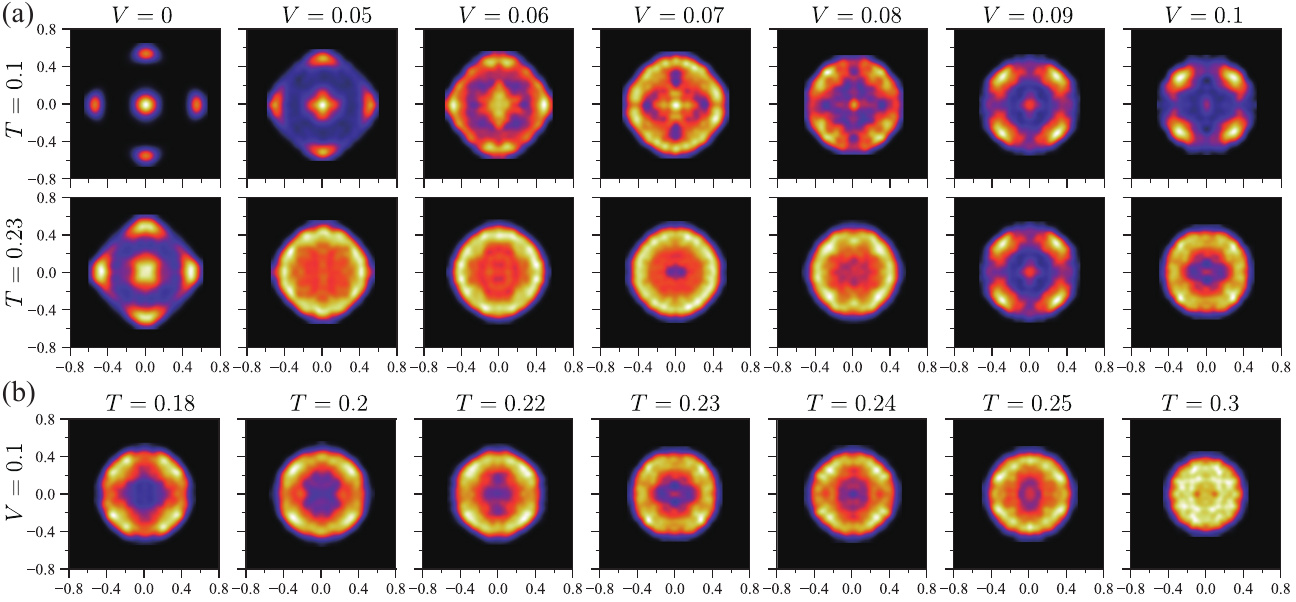}
\caption{Histograms of the vison order parameter $\phi$. The first row, corresponding to $T=0.1$, illustrates the changes in the histogram associated with the first-order transition between the LN phase and the VP phase at finite temperature. This transition features a change from a face-centered cubic anisotropy to a corner-cubic anisotropy for the histogram peaks. The transition point is identified at $V=0.07(2)$ for this temperature. The second row depicts the histogram at $T=0.23$, the LN order appears to melt directly into the disordered phase, with a transition again close to $V=0.07(2)$. These data serve as a supplement to the right column of the  Fig. 4 in the main text. In the third row of the figure, with $V$ fixed at $0.1$, the VP phase disappears as the temperature increases. Based on such results and those with finer temperature and $V$ scans (not shown), we estimate the location of the multicritical point around $(V, T)=(0.07(2), 0.22(2))$ (marked by the red star in Fig. 1 in the main text).}
	\label{fig:figs5}
\end{figure}

\section{Order parameter and histogram in the LN-VP phase transition}
\label{sec:first}
In this section, we present supplementary data regarding the first-order transition between the LN and VP phases, as indicated by the square markers in the overall phase diagram shown in Fig.1 in the main text. The order parameters $\langle D \rangle$ and $|\phi|$ as a function of $V$ at $T=0.1$ for various system sizes are displayed in Fig.~\ref{fig:figs4}. The clear change of the finite size dependence around $V \sim 0$, indicates that $\langle D \rangle$ and $|\phi|$ behave differently in the LN and VP phases. 

To determine the nature of the phase transition, we consider the histograms of the vison order parameter $\phi$, as shown in Fig.~\ref{fig:figs5}. We first focus on $T=0.1$ (first row in Fig.~\ref{fig:figs5}, corresponding to the lower gray dashed line in Fig. 1 in the main text) as an example. We observe  a transition in the histogram peaks from a face-centred cubic structure to a corner-cubic structure, with a coexistence of both at the transition point at $V=0.07$ suggesting a first order phase transition. Meanwhile, we see the LN phase start to vanish at $V=0.05$ and the VP phase appearing at $V=0.09$. Thus, the transition temperature is determined to be $V=0.07(2)$, an estimate which is consistent with the transition point determined from the kinetic energy measurement. This analysis can also be applied to identify transition points at other temperatures. In the second row of Fig.~\ref{fig:figs5}, which displays the histogram at $T=0.23$ (marked by the upper gray dashed line in Fig. 1 in the main text), the LN order appears to melt directly into the disordered phase, with a transition again close to $V=0.07(2)$. The third row of Fig.~\ref{fig:figs5} illustrates the disappearance of the VP order at a fixed value of $V=0.1$ as the temperature increases. This disappearance begins at $T=0.23$, suggesting that the VP phase transitions directly to a disordered phase for this value of $V$. 

\section{Numerical results for the LN-disordered phase transition }
\label{sec:ln}
In this section, we present the raw QMC data for the nematic order parameter and its Binder ratio for the transition between the LN and disordered phases in Fig.~\ref{fig:sln}. This data serves as supplemental information for the left panel of Fig.~\ref{fig:fig2} in the main text. We conduct similar simulations and analyses for other values of $V$ and summarize the transition temperatures in Table.~\ref{tableln}, as illustrated in the phase diagram in Fig.~\ref{fig:fig1} in the main text.

\begin{figure}[htp!]
\centering
\includegraphics[width=\linewidth]{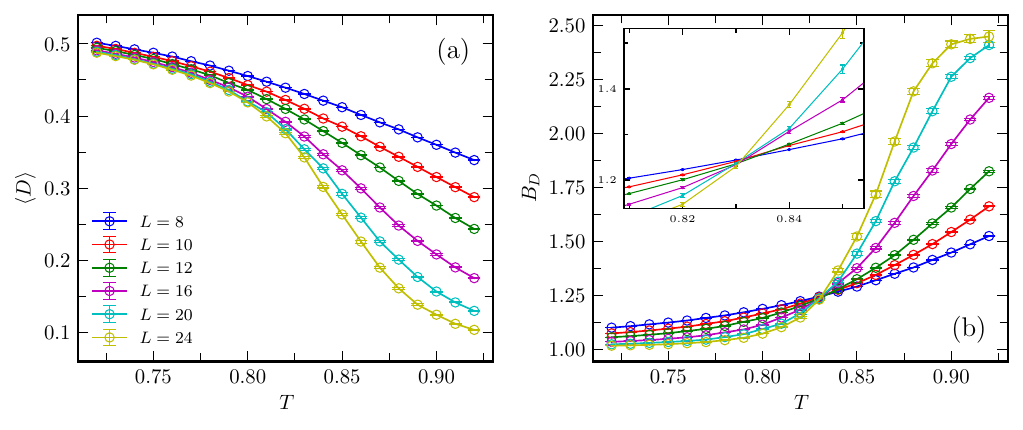}
\caption{ QMC data for (a) the nematic order parameter $\langle D \rangle$ and (b) its Binder ratio as a function of temperature at $V=-0.5$. The crossing points for various system sizes indicates the transition temperature $T_c=0.83(2)$. The errorbars are determined based on the temperature intervals in our simulations. The data collapse using the standard values of the critical exponents from the 3-state Potts universality class is shown in the left panel of Fig.~\ref{fig:fig2} in the main text.  }
	\label{fig:sln}
\end{figure}

\begin{table}[htp!]
    \centering
    \caption{Transition temperatures for various $V$ values in the LN-disordered phase transition.}
    \begin{tabular}{cccccc} 
        \hline
        $V$ & -1 & -0.5  & -0.2  &   -0.1 & 0  \\ \hline

        $T_{c}$              & 1.57(2)  & 0.83(2)           & 0.46(2)      & 0.35(2)      & 0.29(2) \\  \hline
              
    \end{tabular}
    \label{tableln}
\end{table}

\section{Theoretical expectations for the finite-temperature VP-Disorder transition}
\label{sec:SMV}
In this section, we analyse the behaviour of the theory Eq. (2) in the main text at finite temperature. 
We work in $4-\epsilon$ dimensions to study its fixed point and the finite-temperature phase diagram.
For simplicity, let us work with a single scalar first 
\begin{align}
    S=\int d\tau d^d x\, \mathcal{L},\quad {\rm with}\quad  \mathcal{L}=\frac{1}{2}\partial_{\mu}\phi\,\partial^{\mu}\phi+\frac{1}{2}\phi^2+\frac{1}{4!}\lambda \phi^4
\end{align}
To study the physics at scale much larger than inverse temperature, one can expand the scalar theory using Fourier/Matsubara modes along the temporal circle
\begin{align}
    \phi(x,\tau)=\phi_0(x)+\sum_{2n \in \mathbb{N}} \phi_{n}(x) e^{\frac{2\pi n}{\beta} \tau}+\phi^\dagger_{n}(x) e^{-\frac{2\pi n}{\beta} \tau}.
\end{align}
The temporal circle has radius $\frac{\beta}{2\pi}$. 
Notice that due to the $Z_2$ gauge symmetry, the scalar can take both periodic and anti-periodic boundary conditions. The index $n$ for the Matsubara modes can be half integer. After the Fourier expansion, we get 
\begin{align}
    S=\int d^d x \mathcal{L}
\end{align}
with
\begin{align}
    \mathcal{L}=\frac{1}{2}\partial_{\mu}\phi_0\partial^{\mu}\phi_0+\sum_{2n \in \mathbb{N}} \partial_{\mu}\phi_n \partial^{\mu} \phi^\dagger_n+\sum_{2n \in \mathbb{N}} \frac{4\pi^2 n^2}{\beta^2}\phi_n \phi^\dagger_n+ \frac{\lambda }{2}\phi_0^2\sum_{2n \in \mathbb{N}} \phi_n \phi^\dagger_n++\frac{\lambda}{4!}\phi_0^4.
\end{align}
The theory is a massless scalar interacting with infinitely many massive scalars. The one loop diagram with massive loop will renormalize the mass of the zero mode, we get 
\begin{align}
    m^2_{\rm th}=\sum_{2n\in \mathbb{N}}\frac{\lambda}{\beta}\int\frac{d^3k}{(2\pi)^3} \frac{1}{k^2+(\frac{2\pi n}{\beta})^2}=-\frac{\lambda}{2\beta^2}\sum_{2n\in \mathbb{N}} n = -\frac{\lambda}{4\beta^2} \zeta(-1)=\frac{\lambda}{48}\beta^{-2}.
\end{align}
The works of  \cite{PhysRevD.9.3357,PhysRevD.9.3320} studied scalar theories without the $Z_2$ gauge symmetry, and obtained $m^2_{\rm th}=\frac{\lambda}{24} \beta^{-2}$. 
The $Z_2$ gauge symmetry reduces the size of the thermal mass by a factor 2. 
Working in $4-\epsilon$ dimension, we find that the beta function of the $\lambda$ coupling takes the form
\begin{align}
    \beta_{\lambda}=-\epsilon \lambda+\frac{3}{16\pi^2} \lambda^2.
\end{align}
The fixed point at $\lambda=16\pi^2 \epsilon/3$ is the critical Ising model. Substituting the critical coupling $\lambda$ into the thermal mass~\cite{Chai:2020zgq}, we get 
$$ m^2_{\rm th}=\frac{\pi^2\epsilon}{9} \beta^{-2}.$$
At $\epsilon=1$, that is $D=2+1$ dimensions, the above formula will be modified. A simple dimensional analysis, however, tells us that
\begin{align}
    m^2_{\rm thermal}=c T^{1/\nu},
\end{align}
where $c$ is a dimensionless constant.
The finite temperature occurs when $m^2_{\rm thermal}+m^2=0$. Near the quantum critical point,  $m^2 \sim -|V-V_c|$. Together they lead to  
\begin{align}
    T_c\sim |V-V_c|^{\nu}.
\end{align}

Even though we have used the critical Ising model as an example, the above discussion can be applied to other conformal field theories (CFT), such as the Cubic* fixed point in Eq. (2) in the main text. The above discussion suffers from a significant subtlety. Applying it to the Ising* transition~\cite{fradkin1979phase,tupitsyn2010topological,nandkishore2012orthogonal,somoza2021self} from the toric code phase to the trivial phase, one would conclude that the phase transition survives at finite temperature. 
It is, however, well-known that there is rather a cross-over instead of a true phase transition, since the toric code phase does not persist at finite temperature~\cite{castelnovo2007entanglement,nussinov2008autocorrelations,hastings2011topological}. The toric code phase has massive anyons, including the $e$ and $m$ particles, the thermal fluctuation of these particles will destroy the topological order. For our loop model, the local constraint completely forbids the creation of free electrically charged particles, the spinons~\cite{Krishanu2015}. The above discussion focusing solely on the magnetically charged degrees of freedom, the visons $\phi_i$ in~Eq. (2), therefore makes sense.

We now discuss what could be the critical behavior of Eq. (2) when put at finite $T$. We expect a scalar theory respecting the Cubic symmetry in two spatial dimensions. 
One candidate theory that may describe the corresponding VP-Disorder transition is three decoupled copies of Ising models, based on the following reasoning. 
Let us denote the spin operator and the energy operator of the three Ising copies as $\sigma_i$ and $\epsilon_i$. 
The CFT preserves the three $Z_2$ symmetries that flip the signs of $\sigma_i$ as well as the permutation symmetry $S_3$ which permutes the three decoupled Ising copies. 
Notice the Cubic group is isomorphic to $S_3\ltimes(Z_2)^3$. 
The phase transition is triggered by a coupling $s O$ with $O=\epsilon_1+\epsilon_2+\epsilon_3$: when $s>0$, the system is in a disordered phase and when $s<0$, the three $\sigma_i$'s get expectation values (vev) which can be positive or negative, leading to eight corner-cubic states corresponding to the VP phase. 
It is important to check the stability of the decoupled Ising fixed point. The CFT has an marginal operator $O_2=\epsilon_1\epsilon_2+\epsilon_2\epsilon_1+\epsilon_1\epsilon_3$ which has scaling dimension $2$ since $\Delta_{\epsilon}=1$ for the Ising CFT. Turing on the deformation 
\begin{align}
S_{\rm deformation}= K \frac{1}{\sqrt{3}}\int d^2 x \epsilon_1\epsilon_2+\epsilon_2\epsilon_1+\epsilon_1\epsilon_3,
\end{align}
we can calculate the beta function of $K$ using conformal perturbation theory~\cite{PhysRevB.24.6508}
\begin{align}
L \frac{d K}{d L}= \beta(K)= \frac{8}{\pi} K^2+\cdots.
\end{align}
This means that the decoupled Ising fixed point is stable only when $K<0$.
Approximately, $O_2\approx \phi_1^2\phi_2^2+\phi_2^2\phi_3^2+\phi_1^2\phi_3^2$.
One can check that $K<0$ favors the corner cubic phase/VP phase.
Near the QCP, we thus expect a VP-Disorder critical behavior in the (decoupled) Ising universality class. 

We give below further arguments to justify the decoupled Ising CFT using dimension continuation. 
The $\lambda \phi^4$ action Eq. (2) with Cubic symmetry can be studied in $D=4-\hat{\epsilon}$, as was done in the classic paper~\cite{aharony1973critical}.
In addition to the stable Cubic fixed point, one also find the decoupled Ising fixed point which in this case is a tricritical point.
One can also consider the $D=2+\hat{\epsilon}$ expansion. 
From both conformal bootstrap and Borel re-summation~\cite{El-Showk:2013nia}, we know that the operator $\epsilon$ of Ising model has a scaling dimension slightly below $D/2$, when we are in $D=2+\hat{\epsilon}$ dimensions. 
This means that the deformation of the operator $O_2$ is slightly relevant, the beta function then becomes 
\begin{align}
    \beta(K)= \delta K+\frac{8}{\pi} K^2+\cdots,
\end{align}
where $\delta=D-\Delta_{O_2}$ is a small positive number. 
In addition to the decoupled fixed point at $K=0$, we have another fixed point at $K= -\frac{\pi}{8} \delta$. 
So the picture is consistent with each other in both $D=2+\hat{\epsilon}$ and $D=4-\hat{\epsilon}$, the Cubic CFT is stable, and there is a tricritical fixed point given by the decoupled Ising model. 
They merge as $D\rightarrow 2$. 
The appearance of an operator which is exactly marginal at $D=2$ is a sign of fixed points merging~\cite{Gorbenko:2018ncu}.  

The above field theory analysis applies to the physics near the quantum critical point. Away from the QCP, however, another candidate for the VP-Disorder transition is the 4-state Potts universality class. 
The gauge invariant order parameter for the VP phase is given by the kinetic $t_i$-terms, which correspond to scalar bi-linear fields  $t_{1}=\phi_1\phi_2, t_{2}=\phi_2\phi_3, t_{3}=-\phi_1\phi_3$~\cite{ran2024hidden,Ran2024Cubic}. The operators $\{\phi_1\phi_2,\phi_2\phi_3,-\phi_1\phi_3 \}$ form a irreducible triplet representation of the Cubic group. 
It is, however, not a faithful representation. The subgroup faithfully represented on the triplet is the permutation group $S_4$, as hinted by the tetrahedron shapes in Fig. 3 (a) and (d) in the main text when $\phi_{i}=\pm {\rm vev}$. 
The breaking of the $S_4$ symmetry may thus lead to the 4-state Potts universality class.

The two CFTs, the decoupled Ising model and the 4-state Potts model, correspond precisely to the two different scenarios of whether fractionalization happens. Near the VP-Disorder transition temperature, the $t$ term should be identified as 
$\{\phi_1\phi_2,\phi_2\phi_3,-\phi_1 \phi_3\}$. 
If ``fractionalization'' does not happen, the candidate CFT is the 4-state Potts model and the corresponding global symmetry is $S_4$.
The $t$ terms $\{\phi_1\phi_2,\phi_2\phi_3,-\phi_1 \phi_3\}$ and the $\phi_i$ in fact are in the same irreducible representation of $S_4$. 
To be more precise, the group $S_4$ has a three dimensional representation usually called the ``standard'' representation, and $\phi_i$ is in this representation. 
The ``standard'' representation of $S_4$ has a rank-3 symmetric tensor $d_{ijk}$, which can be constructed using the recursive procedure discussed in~\cite{RKPZia_1975}. The bilinear fields $d_{ijk}\phi_j\phi_k$ are therefore also in the ``standard'' representation. 
If one matches the convention, $d_{ijk}\phi_j\phi_k$ are precisely $\{\phi_1\phi_2,\phi_2\phi_3,-\phi_1 \phi_3\}$. 
Lattice observables are classified by the symmetry of the CFT. Since the $\phi_i$ field and the $t$ term fields $\{\phi_1\phi_2,\phi_2\phi_3,-\phi_1 \phi_3\}$ transform in the same irrep of the  $S_4$ group, they should have the same scaling dimensions, as a consequence of universality. 
The non-fractionalized scenario above is very similar to the LP-Disorder transition where the $\phi_i$ and dimer have the same scaling dimension.
If fractionalization indeed happens, the $\phi_i$ field and the $t$-terms can have different scaling behaviors. 
We have shown numerically that at the lattice size available to us, the $\phi_i$ field and the $t$ terms have different scaling dimensions therefore indicating ``fractionalization'' happens at the current lattice sizes scale.
The remaining unresolved ambiguity is about when the size of the lattice is large enough for us to neglect the influence of the quantum critical point. If ``fractionalization'' does not survive, we should get the 4-state Potts model universality class. If ``fractionalization'' survives, we should have the decoupled Ising universality class.
We note that both CFTs have marginal operators. This is well known for the 4-state Potts model~\cite{PhysRevLett.44.837,PhysRevB.22.2560}. 
For the decoupled Ising universality class, the marginal operator is $O_2$ which we just discussed. 
The logarithm corrections induced by these marginal operators may make it difficult to distinguish the two CFTs on finite lattice sizes. 

\end{widetext}

\end{document}